\documentclass[fleqn,10pt]{SelfArx} 

\usepackage[english]{babel} 

\usepackage{cmap} 

\definecolor{color1}{RGB}{0,0,90} 
\definecolor{color2}{RGB}{0,20,20} 

\usepackage{hyperref} 

\hypersetup{
	hidelinks,
	colorlinks,
	breaklinks=true,
	urlcolor=color2,
	citecolor=color1,
	linkcolor=color1,
	bookmarksopen=false,
	pdftitle={Title},
	pdfauthor={Author},
}

\JournalInfo{Journal, Fluid Dynamics, 2024} 
\Archive{Preprint} 

\PaperTitle{Unstable gas flow in a flat channel under the influence of a transverse force field: self-oscillations of the jet} 

\Authors{S.D. Korolkov\textsuperscript{1, 2}*, V.V. Izmodenov\textsuperscript{1, 2, 3}} 
\affiliation{\textsuperscript{1}\textit{Space Research Institute of RAS, Moscow, 117485 Russia}} 
\affiliation{\textsuperscript{2}\textit{Moscow center for fundamental and applied mathematics, Lomonosov Moscow State University,  Moscow 119991, Russia}} 
\affiliation{\textsuperscript{3}\textit{Ishlinsky Institute for Problems in Mechanics of RAS, Moscow, 119526 Russia}} 
\affiliation{*\textbf{Corresponding author}: sergey.korolkov@cosmos.ru} 

\newcommand{\keywordname}{Keywords} 

\Abstract{A subsonic flow of an ideal gas through a flat channel is considered. In a finite channel region, a transverse force field acts, attracting the flow toward the channel axis. This field leads to flow instability, even at a relatively weak magnitude. For a weak force field, the instability is symmetrical relative to the channel axis, while for a strong force field, it becomes asymmetrical and self-oscillations occur. The characteristics of this instability have been studied numerically in this work. Possible astrophysical implications of this phenomenon are discussed.}

\begin{document}

\maketitle 

\tableofcontents 

\thispagestyle{empty} 


\section*{INTRODUCTION} 

\addcontentsline{toc}{section}{INTRODUCTION} 

\begin{figure}
	\center{\includegraphics[width=1.0\linewidth]{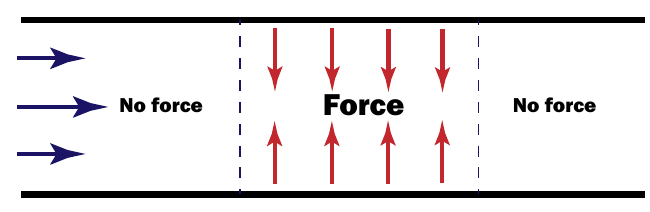}}
	\caption{Schematic illustration of the flow in a flat channel under the influence of a force field. The force field is represented by red arrows that act only in a specific part of the channel.}
	\label{A1}
\end{figure}


The problem discussed in this paper arises from astrophysical applications. According to Parker model \cite{Parker1958, Parker1961} the magnetic field of a star is frozen into the stellar wind plasma flow and has a predominant azimuthal component due to stellar rotation. In a supersonic stellar wind, the magnetic force is insignificant compared to the dynamic pressure of the plasma (the Alfven Mach number is significantly greater than 1). However, after the stellar wind termination shock \cite{izmod2006} at which the stellar wind becomes subsonic, the magnetic force becomes substantial and the azimuthal magnetic field starts to press the plasma flow towards the rotation axis of the star \cite{1974Yu}. Consequently, if the interstellar medium is at rest or slow enough, it leads to the formation of two plasma jets directed along the rotation axis of the star \cite{Golikov2017MNRAS, Korolkov2021, Korolkov2023}. This process can be compared to the confinement of plasma by an azimuthal magnetic field in a tokamak. Thus, two plasma jets form a tube-like structure of the astrosphere, which differs significantly from the classical astrosphere/heliosphere \cite{Baranov1970, Baranov1993}.

Preliminary results (which will be published in future papers) show that the astrophysical jet may be unstable for the strong stellar magnetic field. In this paper, we propose a mechanism responsible for the instability. To perform a more effective study we consider a simplified two-dimensional model. The plasma flow in the astrophysical jet resembles gas flow through a pipe. However, modeling the asymmetric modes in an axisymmetrical geometry of the flow in a pipe would require three-dimensional calculations. To stay in 2D we consider a flat channel i.e. the flow between two parallel planes. To mimic the magnetic field we introduce a body force that is directed toward the middle (or central) plane of the channel. The magnitude and shape of this force are varied. The paper considers both discontinuous and continuous fields of the force. Further, the channel will be assumed to be two-dimensional, the central plane is called the channel's axis of symmetry.

Since the magnetic force does not affect the plasma immediately behind the termination shock, but at a certain distance from it, the body force also acts not immediately at the beginning of the channel, but at some distance from it (see Figure~\ref{A1}). Also, the force field is limited to a finite region, although it is quite large. This is done to reduce the computational difficulty of specifying the boundary conditions at the end of the channel. We will discuss this in more detail in the next section.


It is worth noting that the instability of plasma jets in the astrosphere was previously discussed \cite{Opher2015}. The authors associate this instability with the influence of hydrogen atoms, which act as an effective force across the jet and lead to Rayleigh-Taylor instability. However, in the present work, a different type of instability will be demonstrated that occurs even in the absence of atomic neutrals.


The work structure is as follows: in Section~\ref{model} the mathematical model, boundary conditions, and numerical method are described, and the problem is formulated in dimensionless form; Section~\ref{sec: results} shows the results of calculations, which are presented separately for discontinuous ``strong'', ``weak'' and ``medium'' force fields as well as for continuous field. In the last section, the results are summarized and discussed.


\section{MODEL} \label{model}

In this paper the simple model is considered, aiming to explore the instability caused by the specifically introduced force. Figure~\ref{A1} schematically illustrates the geometry of the considered problem. We consider a two-dimensional planar flow in a channel where a force acts perpendicular to the flow and is directed toward the channel axis. The gas is assumed to be ideal with a constant ratio of heat capacities ($\gamma = 5/3$). Thus, this problem will be solved within the framework of pure gas dynamics by using the Euler equations. It is also assumed that there is a body force (the term in the motion equation equal to $\rho F$). The system of equations in a divergent form that is convenient for numerical solution is shown below:

\begin{align}
\begin{cases}
\dfrac{\partial \rho}{\partial t} + \textbf{div}( \rho{\bf V}) = 0,\\[3mm]
\dfrac{\partial (\rho {\bf{V}})}{\partial t} + \textbf{div}( \rho{\bf V} {\bf V} + p\hat{\bf I}) = \bf Q_2,\\[3mm]
\dfrac{\partial E}{\partial t} + \textbf{div}( (E+p) {\bf V}) = {\bf Q_2} \cdot  {\bf V},\\[3mm]
\end{cases}
\label{sys1}
\end{align}

where
$
E = \frac{p}{\gamma-1}+ \frac{\rho V^2}{2}
$ is the total energy density; $\rho,\ p,\  {\bf{V}}$ are the gas density, pressure and velocity, respectively. The energy source in the third equation of System~\ref{sys1} (the equation for the total energy of the system) is the scalar product of the momentum source and velocity. The equation of state for a gas is $p = n k_B T$, where $n$ and $T$ are the gas number density and temperature, respectively; $k_B$ is the Boltzmann constant. 

The momentum source, ${\bf Q_2}$, represents the body force acting on the flow. The formula for calculating the momentum source will be given immediately in dimensionless form in the next section.

\subsection{Boundary conditions and dimensionless formulation} \label{bound}

The walls of the channel are assumed to be rigid and impermeable, therefore $V_n = 0$. The incoming flow is specified at the left boundary of the channel. Soft boundary conditions (see Subsection~1.2) are applied at the right boundary.

Let us assume the channel width as the unit of length, and the density and velocity of the incoming flow as the units of density and velocity respectively. Thus, the considered problem depends on {\bf the Mach number of the incoming flow} ($M_0$) and {\bf the magnitude of the force} ($F$, see Equation~\ref{Q2}). The width of the area of action of the force is assumed to be 10 dimensionless units. The dimensionless pressure of the incoming flow is determined as follows: $\hat{p} = \dfrac{1}{\gamma M_0^2}$. The Mach number of the incoming flow, $M_0$, in all calculations is assumed to be 0.29.

Since the channel cannot be infinitely long in the calculations, it was made as long as possible, with a length of 80 dimensionless units ($0 < X < 80$). The region where the force field operates is located in the center of the channel, at $35 < X < 45$. The channel height varies from 0 to 1 ($Y = 0.5$ is the channel axis).

The formula for calculating the momentum source, ${\bf Q_2}$, is as follows (all quantities are assumed to be dimensionless):

\begin{align}
Q_{2 x}(x, y) & =  0, \nonumber \\[2mm]
Q_{2 y}(x, y) & = 
\begin{cases}
=   -\rho\  F\  \mathrm{sign}(y - 0.5),\\ \ \ \ \ \ \ \ \mathrm{for} \ \ 35 < x < 45;\\[1.5mm]
= 0 ,\ \mathrm{otherwise}.
\end{cases}
\label{Q2}
\end{align}

\subsection{Numerical approach} \label{num}

Due to the geometry of the problem, a uniform Cartesian grid was used. The grid resolution may vary for different calculations, so it will always be specified separately. The maximum resolution was 32,000 x 400 computational cells (since the channel is very extended along the X-coordinate). The program is implemented for GPU processors using parallel programming technology, CUDA C++.
For calculations, Godunov-type finite volume methods (HLLC solvers, see \cite{Miyoshi2005}) are used. We also tested the method with an exact solver of the Riemann problem \cite{godunov1976}, and the results were consistent with the HLLC solver. HLLC was used in this study to speed up calculations. The minmod limiter was also used to obtain a second-order spatially accurate solution (since there are no discontinuities in gas-dynamic variables in the flow).
On the left boundary of the computational domain, the values of gas-dynamic quantities were fixed. On the right boundary, soft boundary conditions are used; these conditions require that the derivatives of the gas-dynamic quantities with respect to the normal direction of the boundary (along the X-axis) are zero.

To diminish the effect of the boundary conditions at the entrance and exit of the channel (the reflection of incoming waves), the following method was used: at the beginning ($0< X < 20$) and at the end ($60 < X < 80$) of the channel the first-order computational scheme and the HLL solver has been employed \cite{Miyoshi2005}. Such a scheme has a much larger numerical viscosity as compared with the ``HLLC+minmod'' scheme and reduces the impact of reflected waves on the flow.

\begin{figure*}[ht]\centering
	\includegraphics[width=1.0\hsize]{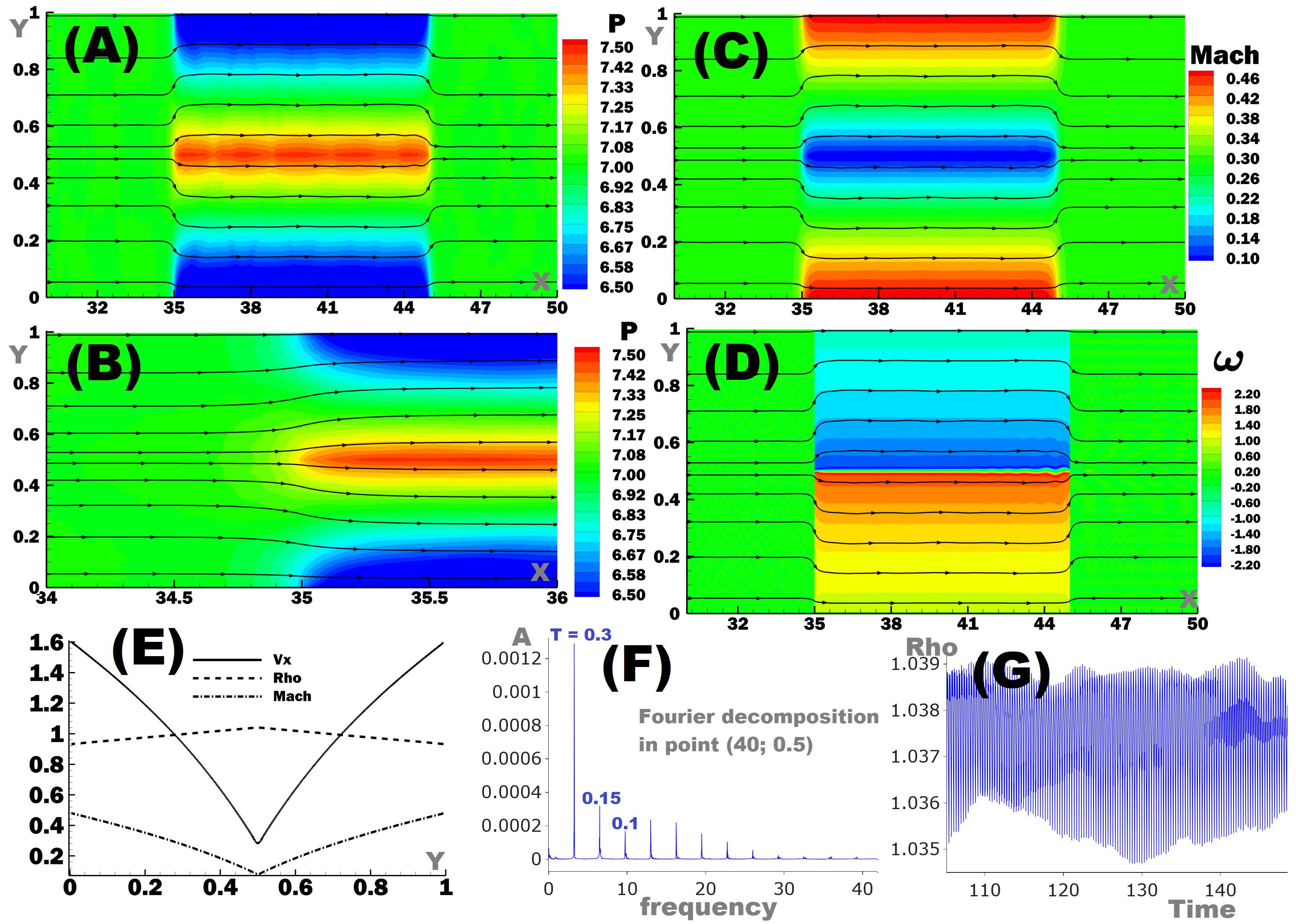}
	\caption{The snapshot with the streamlines and pressure isolines (A, and B for a closer view),  the Mach number (C), the vorticity isolines (D), the X-velocity, density, and Mach number profiles across the channel at $X = 40$ (E), the density dependence on time at a certain point in the flow (G) and its Fourier decomposition (F). $F = 2.5$ and $M_0 = 0.29$. See "Video-1.avi" in Data availability.}
	\label{A2}
\end{figure*}

The initial conditions will not affect the solution at large times, but at the initial time moment, we set a uniform flow. That is, it is assumed that the force field starts to act from the initial moment in time.

It should be noted that the formulated above initial and boundary conditions are symmetrical with respect to the channel axis, and the numerical scheme is also symmetrical. As a result, obtaining asymmetric solutions is not possible. To generate asymmetric solutions, small (1\%  of magnitude) periodic density disturbances at the left boundary condition for the lower part of the channel were introduced for a short period ($0 < t < 3$) and then switched off. This approach leads to the emergence of asymmetric disturbances in the channel that may either disappear if the symmetric solution is more stable or lead to an asymmetric global solution.

\section{RESULTS} \label{sec: results}

In this section, the main results of our calculations with $M_0 = 0.29$ ($\hat{p} = 7$) will be discussed. To make it easier to follow, the section is divided into four subsections. The first three subsections show results with a discontinuous force field, as defined by Equation~\ref{Q2}. The first subsection presents the results for a weak force field ($F = 2.5$). The second subsection shows the results for a strong force field ($F = 4$); the third subsection presents the case with an average force field value ($F = 3$). This division is quite rough and is made to demonstrate different flow regimes are obtained. The last subsection presents additional results using the continuously differentiable force field defined by Equation~\ref{Q22}.

\subsection{Weak force field}
\label{weak}

Figure~\ref{A2} shows the results of the calculation for $F = 2.5$. The computational grid size is 32,000 x 400 cells. Panel (A) of the figure shows the snapshot with the streamlines and pressure isolines. It should be noted that the scales of the X and Y axes are different to show the entire domain of the force field. Panel (B) shows a closer view of the initial cross-section of the force action. In this case, the scales of the axes are the same.

The pattern of the obtained flow can be explained as follows: the pressure gradient tries to compensate for the force action, so the pressure profile reaches a maximum at the channel axis and decreases towards the walls (Figure~\ref{A2}~A,~B). In general, the pressure change is quite small and is in the range of $-10, +7\%$ of the incoming flow pressure. Since there is no thermal energy influx (let us recall that the source term appears in the third equation of System~\ref{sys1} because this is a total energy equation; if the internal energy equation had been written, there would not have been any sources), entropy in the particles must be conserved, so the density profile qualitatively repeats the pressure profile, reaching a maximum at channel axis (density isolines are not shown in the figure, however one-dimensional profiles are presented in Panel E, see description below). Also, note that the formation of a pressure maximum at the channel axis leads to the emergence of a gradient of pressure along the channel near the initial cross-section ($X=35$). This gradient hinders the flow, causing the streamlines to flow around an obstacle, which can be clearly seen in Panel B.

The presented solution is not fully stationary but rather quasi-periodic. Instabilities occur in the flow and there are small oscillations in the pressure profile in Panel A. The flow pattern is shown at a specific point in time (snapshot) long after the start of the calculation. The non-stationary behavior of the flow can be explored in the video (see ``Video-1.avi'' in data availability section). It should be additionally noted that this instability is completely symmetrical relative to the channel axis.

Panels C and D in Figure~\ref{A2} show the Mach number and vorticity isolines ($\frac{1}{2}\mathrm{rot}({\bf V})$). Note that there are no large-scale vortexes in the flow. The vorticity field is discontinuous because the velocity derivatives are discontinuous due to the influence of a discontinuous force field. Near the end of the force action region ($X = 45$), flow oscillations on the channel axis can also be seen in Panel D. Panel E shows the X-velocity, density, and Mach number profiles across the channel at $X = 40$.  Here it is seen that the flow is faster near the walls, which explains the vorticity field in Panel D.

Panel G shows density versus time at $X=40,\ Y = 0.5$. This point is located in the middle of the force field region. Panel F displays this density as a Fourier series. The $X$-axis represents the frequency of the oscillation, which is equal to 1 divided by the period of the oscillation. The $Y$-axis shows the values of the coefficients in the Fourier expansion of the density. Panel F displays the fundamental period, $\mathrm{T}=0.3$, and its multiples. In this case, the solution can be considered quasi-periodic with a fundamental period of $0.3$. This period is associated with the time it takes for disturbances to pass from one wall of the channel to the other. For the incoming flow, this period should be equal to the Mach number ($M_0 = 0.293$), but for the disturbed flow it will be slightly larger ($\approx 0.296$) due to the change in the speed of sound across the channel width, which coincides well with the frequency observed in the numerical solution. It should be noted that if the grid resolution changes, the harmonic pattern remains completely intact (these results are not shown here).

\subsection{Strong force field}\label{strong}

\begin{figure*}[ht]\centering
	\includegraphics[width=1.0\hsize]{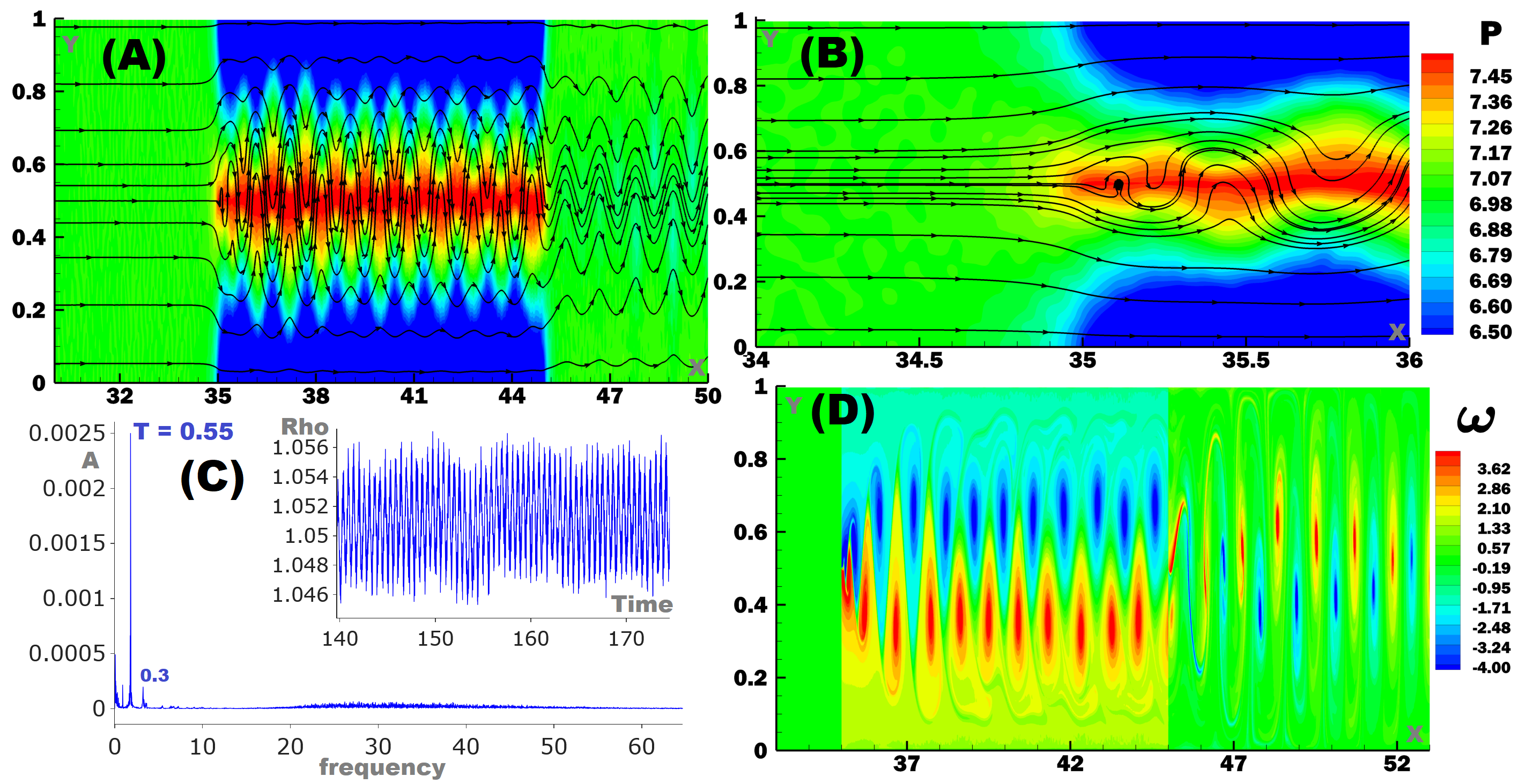}
	\caption{The snapshot with the streamlines and pressure isolines (A, and B for a closer view),  the vorticity isolines (D), the density dependence on time at ($X = 40,\ Y = 0.5$) and its Fourier decomposition (C). $F = 4$ and $M_0 = 0.29$.  See "Video-2.avi" in Data availability.}
	\label{A3}
\end{figure*}

Figure~\ref{A3} shows a snapshot of the solution for $F = 4$. The computational grid size is 32,000 x 400 cells. The flow in this case is qualitatively different from the considered above. Panel A shows the streamlines and pressure contours for the flow, while Panel B shows the same information but with a closer view of the region near $X = 35$. Strong asymmetric oscillations in the flow are observed (see ``Video-2.avi'' in data availability section). A vortex forms near point (35, 0.5) in panel B. The gas flows around the vortex and something similar to a Karman vortex street forms behind it. Vorticity isolines ($\frac{1}{2}\mathrm{rot}({\bf V})$) are shown in Panel D. The entire system of formed vortices is clearly visible.

Plots of the Fourier series expansion of the density and the density versus time plots are shown in panel C. Unlike Subsection~2.1, there is no set of multiple periods. Instead, a single period $\mathrm{T} = 0.55$ stands out clearly ($T = 0.3$ also remains noticeable). A set of high-frequency oscillations with frequencies from 20 to 40 with an almost continuous spectrum is also clearly visible. When changing the resolution, the harmonic pattern remains qualitatively preserved (these results are not shown here), although it is not so ideal preservations as obtained in Subsection~2.1. Namely, the main period persists, but the finer the grid resolution is, the more continuous the spectrum of high-frequency disturbances.

Although the exact mechanism of stability loss is still not fully understood. However, in our opinion, the self-oscillation appeared due to a feedback mechanism of the waves reflected from the channel walls. In other words, if a jet is deflected upwards, the pressure in the upper part of the channel increases, which leads to the deflection of the jet downwards, and so on. This mechanism is somewhat similar to the mechanism of oscillation of a submerged jet, as described in \cite{Karlikov1998, Karlikov2009}.

\subsection{Average force field}\label{average}

\begin{figure*}[ht]\centering
	\includegraphics[width=1.0\hsize]{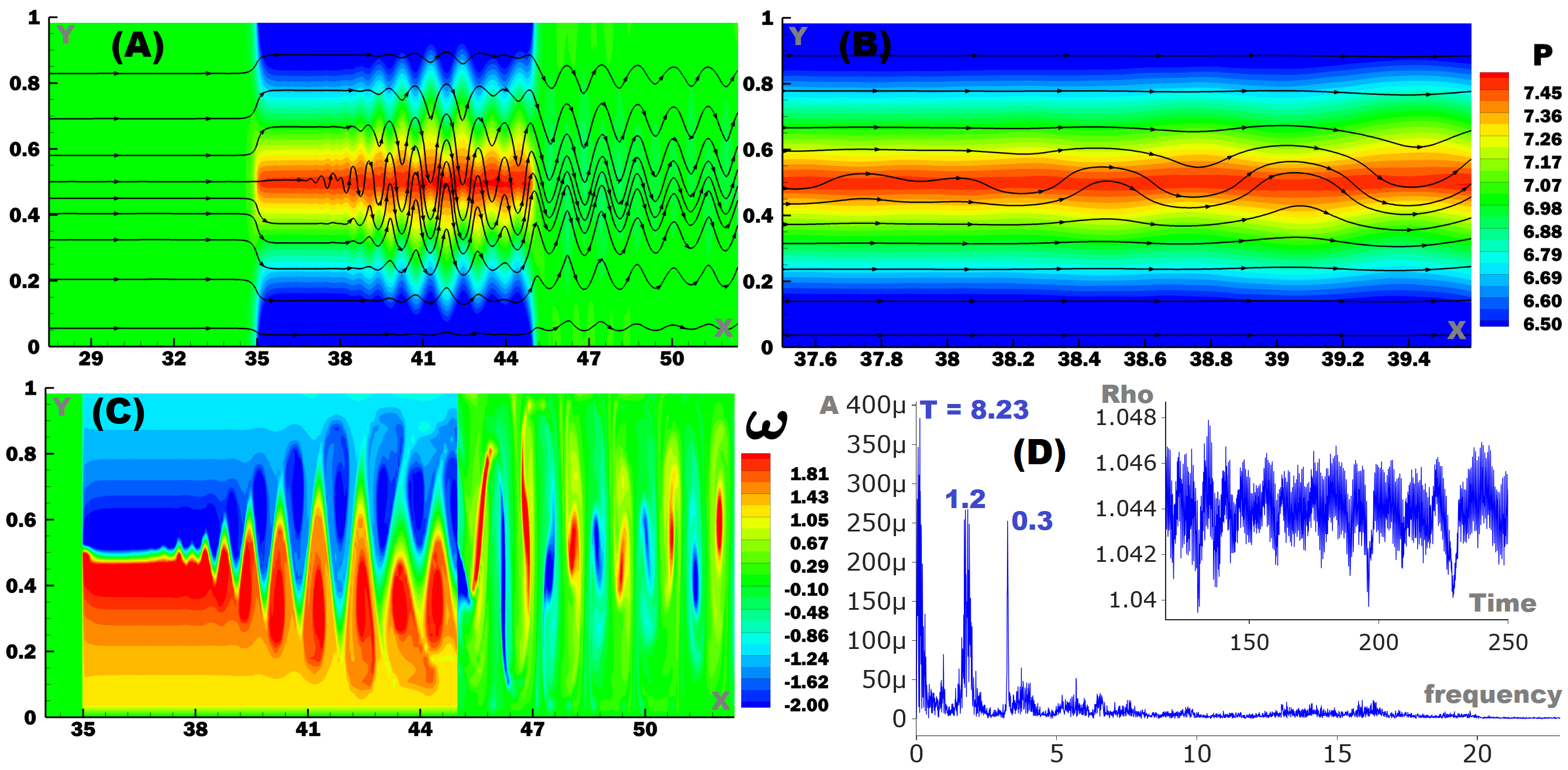}
	\caption{The snapshot with the streamlines and pressure isolines (A, and B for a closer view),  the vorticity isolines (C), the density dependence on time at ($X = 40,\ Y = 0.5$) and its Fourier decomposition (D). $F = 3$ and $M_0 = 0.29$.  See "Video-3.avi" in Data availability.}
	\label{A4}
\end{figure*}

Based on the information presented in the previous sections, it can be concluded that the behavior of the instability of the flow depends on the magnitude of the force field. Specifically, for weak forces (as discussed in Subsection~2.1), the instability is symmetric about the channel axis and mainly consists of a single dominant frequency and its multiple harmonics. In contrast, for strong forces (discussed in Subsection~2.2), the instability becomes asymmetric and includes several low-frequency oscillations, with one prominent oscillation, and a series of higher-frequency oscillations with a nearly continuous spectrum. This section presents a solution involving an intermediate force magnitude ($F = 3$), in which both types of instability are observed simultaneously, transitioning from one to the other.

Figure~\ref{A4} (and ``Video-3.avi", see data availability section) shows the results of numerical simulations of the flow at $F = 3$. The computational grid size is 16,000 x 200 cells. Panels A and B show pressure isolines and streamlines. In this case, instability is visible mainly in the right part of the force zone. In the left part, the flow appears to be symmetrical. However, panel C, which shows vorticity, reveals that the flow is not symmetrical even on the left side. This is expected, as in subsonic flows asymmetric disturbances travel from the right to the left side of the computational domain. This solution allows us to trace the development of asymmetric instability. Fourier spectra in Panel D show a wide range of oscillation frequencies, including low frequencies. However, oscillations with a period of 0.3 persist from weak fields (Subsection~2.1, Figure~\ref{A2}) to strong fields (Subsection~2.2, Figure~\ref{A3}).

It should be noted that the instability appears to be convective. The disturbances grow in space, rather than in time. The flow is not destroyed, and the disturbances are carried beyond the calculated region.

\subsection{Continuous force field}\label{cff}

\begin{figure*}[ht]\centering
	\includegraphics[width=1.0\hsize]{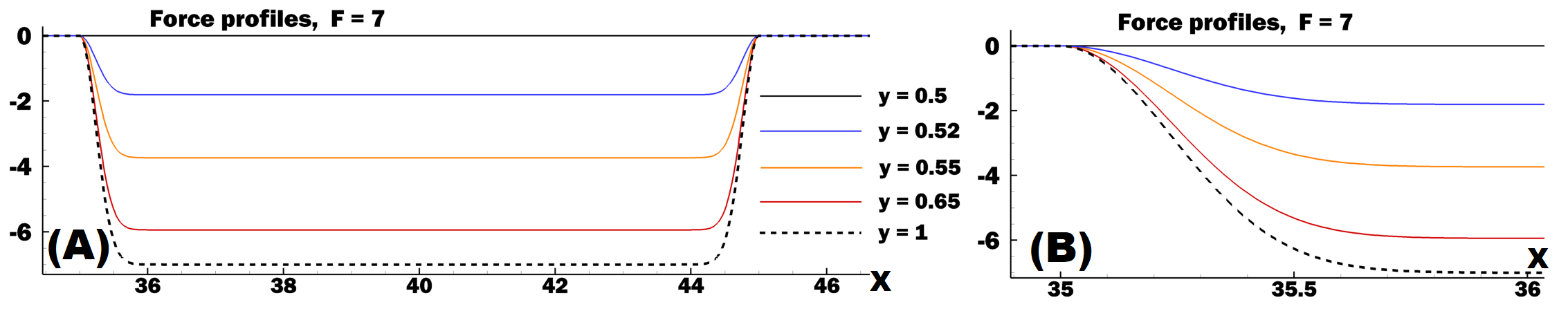}
	\caption{Continuous force field profiles given by Equation~2.4.1 for $F = 7$ (A, and B for a closer view).}
	\label{A5}
\end{figure*}

\begin{figure*}[ht]\centering
	\includegraphics[width=1.0\hsize]{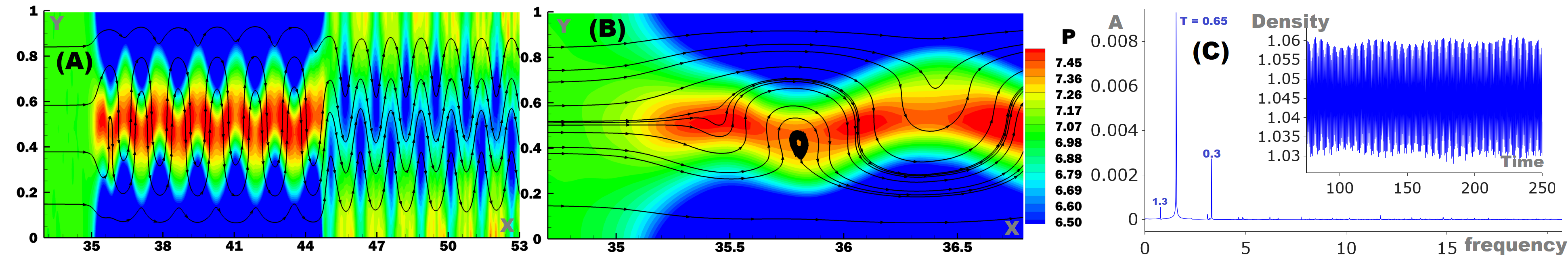}
	\caption{The snapshot with the streamlines and pressure isolines (A, and B for a closer view), the density dependence on time at ($X = 40,\ Y = 0.5$) and its Fourier decomposition (C). $F = 7$, a continuous field is given by the Equation~2.4.1  and $M_0 = 0.29$.  See "Video-4.avi" in Data availability.}
	\label{A6}
\end{figure*}

In this section, the question of whether a discontinuous force field plays a main role in the occurrence of instability is discussed. Some additional calculations were conducted by using a force field which is described by:

\begin{align}
Q_{2 y}(x, y) = 
\begin{cases}
-\rho \ F \  \dfrac{\mathrm{atan}(20(y - 0.5))}{\mathrm{atan}(10)} \times \\[2mm] \ \ \ \ \ \ \ \times  (1 -  \mathrm{exp}(-(3 (x - 35))^2) \times \\[3mm]
\ \ \ \ \ \ \ \times(1 -  \mathrm{exp}(-(3 (x - 45))^2) ,\\[2mm] \mathrm{for}\ \  35 < x < 45,\\[3mm]
0,\ \mathrm{otherwise}.
\end{cases}
\label{Q22}
\end{align}

The profiles of this field for $F = 7$ are shown in Figure~\ref{A5} ($\rho = 1$ for demonstration). The function in the right part of (\ref{Q22}) is continuously differentiable in the entire domain. 

The results of the calculations with $F = 7$ are shown in Figure~\ref{A6} and ``Video-4.avi" (see data availability section). The computational grid size is 16,000 x 200 cells. Asymmetric instability occurs in the flow, similar to what happens with a strong discontinuous force (see Subsection~2.2). The spectrum of the Fourier expansion is shown in Panel C, and it should be noted that the solution in this case is strictly periodic. Two different oscillation frequencies are clearly visible, and neither of them is a multiple of the other. Interestingly, the amplitudes of the other frequencies are negligibly small.


\section*{CONCLUSIONS}

\addcontentsline{toc}{section}{CONCLUSIONS}

In this paper, the problem of the stability of an ideal fluid flow in a flat channel under the influence of a transverse field of forces that pushes the gas toward the channel axis is solved. The main results of this work are summarized as follows:

\begin{itemize}

\item The action of the transverse force field results in an increase of the pressure at the symmetry axis of the channel, as the pressure field attempts to counteract the force. At the interface between the free force region and the region where the force field acts, a pressure gradient forms along the axis of the channel, which hinders the flow. Fluid particles flow around the high-pressure zone. This creates a flow pattern similar to that around a body. It is  believed that the high-pressure area in the flow is responsible for the instability.

\item For weak forces ($F \leq 2.5$), a weak symmetrical instability in the flow forms, which manifests itself as oscillatory flows with a dominant period of $\sim$ 0.3 and all its multiples. This period, to a sufficient degree of accuracy, is the time it takes for the disturbance to travel from one wall of the channel to the other (for an incoming flow, this time would be equal to the Mach number $M_0$).

\item For strong forces ($F \geq 4$), a strong asymmetric instability forms, manifesting itself in auto-oscillations of the flow. Several low-frequency disturbances (including an oscillation with a frequency of $\sim$ 0.3 that persists but is not the dominant one) as well as a large number of high-frequency disturbances with a continuous spectrum emerge in this case. As the force continues to increase, the high-frequency disturbances fade away, leaving only two dominant periods of oscillation (including 0.3). In this case, the feedback mechanism is the reflection of disturbances from the channel walls. This mechanism is responsible for the self-oscillation of the flow.

\item For forces with an average magnitude ($F \approx 3$), it is possible to track the evolution of instability along the channel. This is possible because the time of instability growth roughly corresponds to the time it takes for the gas to pass through the area affected by the force field.

\item An additional study has shown that a force field does not need to be discontinuous in order to create flow instability. It is possible to create instability even with the action of a continuously differentiable force field. The factor is the formation of an increased pressure region near the axis of the channel, which forms a pressure gradient along of X-axis direction and hinders the flow. For example, a linear force of attraction towards the channel axis $\left(F \sim \mathrm{sign}(0.5 - y) \right)$ was found to be insufficient for creating the desired pressure gradient and inducing instability.

\end{itemize}

The study of this instability and its application to astrophysical jets will be continued in future research.


\phantomsection
\section*{Acknowledgments} 

\addcontentsline{toc}{section}{Acknowledgments} 

SK thanks the Theoretical Physics and Mathematics Advancement Foundation `BASIS' grant 22-8-3-42-1.

The research is carried out using the equipment of the shared research facilities of HPC computing resources at Lomonosov Moscow State University  \cite{Voevod2019}

\section*{Data availability}
\addcontentsline{toc}{section}{Data availability} 

The videos released with this paper have been posted on Zenodo under a Creative Commons Attribution license at DOI: \href{https://doi.org/10.5281/zenodo.13992316}{https://doi.org/10.5281/zenodo.13992316}.

The other data underlying this article will be shared on reasonable request to the corresponding author.


\phantomsection
\bibliographystyle{unsrt}
\bibliography{article_3.bib}


\end{document}